\documentclass[iop]{emulateapj}
\usepackage{apjfonts}

\usepackage[tight]{subfigure}
\usepackage{amsmath}
\usepackage{graphicx}
\usepackage{wasysym}
\usepackage{color}
\usepackage{verbatim}
\usepackage{hyperref}
\usepackage{xcolor}
\hypersetup{
    colorlinks,
    linkcolor={red!50!red},
    citecolor={blue!50!blue},
    urlcolor={blue!80!black}
}

\shorttitle{Blackhole feedback and anisotropic thermal conduction}
\shortauthors{Kannan et al.}

\begin{document}

\title{Increasing blackhole feedback induced quenching with anisotropic thermal conduction}

\author{Rahul Kannan\altaffilmark{1}, Mark Vogelsberger\altaffilmark{1,6}, Christoph Pfrommer\altaffilmark{2}, Rainer Weinberger\altaffilmark{2} \\ Volker Springel\altaffilmark{2,3}, Lars Hernquist\altaffilmark{4}, Ewald Puchwein\altaffilmark{5}, R\"udiger Pakmor\altaffilmark{2}}
\email{kannanr@mit.edu}

\altaffiltext{1}{Department of Physics, Kavli Institute for Astrophysics $\&$ Space Research, Massachusetts Institute of Technology, Cambridge 02139, MA, USA} 
\altaffiltext{2}{Heidelberg Institute for Theoretical Studies, Schloss-Wolfsbrunnenweg 35, D-69118 Heidelberg, Germany}
\altaffiltext{3}{Zentrum f\"ur Astronomie der Universit\"at Heidelberg, ARI, M\"onchhofstr. 12-14, D-69120 Heidelberg, Germany}
\altaffiltext{4}{Harvard-Smithsonian Center for Astrophysics, 60 Garden Street, Cambridge, MA 02138, USA}
\altaffiltext{5}{Institute of Astronomy and Kavli Institute for Cosmology, University of Cambridge, Madingley Road, Cambridge, CB3 0HA, UK}
\altaffiltext{6}{Alfred P. Sloan Fellow}

\begin{abstract} 
  Feedback from central supermassive blackholes is often invoked to
  explain the low star formation rates in massive galaxies at the
  centers of galaxy clusters. However, the detailed physics of the
  coupling of the injected feedback energy with the intracluster
  medium is still unclear. Using high-resolution magnetohydrodynamic
  cosmological simulations of galaxy cluster formation, we investigate
  the role of anisotropic thermal conduction in shaping the
  thermodynamic structure of clusters, and, in particular, in
  modifying the impact of black hole feedback. Stratified anisotropically
  conducting plasmas are formally always unstable, and thus more prone
  to mixing, an expectation borne out by our results. The increased mixing efficiently isotropizes the injected
  feedback energy which in turn significantly improves the coupling between the
feedback energy and the intracluster medium. This facilitates an earlier disruption of the
  cool core, reduces the star formation rate
  by more than an order of magnitude, and results in earlier quenching despite an overall lower amount of feedback energy 
  injected into the cluster core. With conduction, the metallicity
  gradients and dispersions are lowered, aligning them better with
  observational constraints. These results highlight the important
  role of thermal conduction in establishing and maintaining
  quiescence of massive galaxies.
\end{abstract}


\section{Introduction}
\label{sec:intro}
Feedback from active galactic nuclei (AGN) has widely been invoked to
explain the quenching and quiescence of massive
galaxies~\citep{Croton2006, Sijacki2007, Booth2009, Choi2012, Li2014,
  Weinberger2016}.  However, the details of how this feedback energy
couples to the surrounding gas are still not properly understood, so
the modelling efforts have been necessarily crude. Despite this
limitation, recent cosmological models have had reasonable success in
regulating the properties of massive central
galaxies~\citep[e.g.][]{Genel2014, Vogelsberger2014, Schaye2015, Sijacki2015, Weinberger2016}.

However, these galaxy formation simulations did not account for
important physical processes related to thermal conduction and
magnetic fields, which can significantly affect the properties of the
intracluster medium (ICM)~\citep{Balbus2000, Carilli2002,
  Quataert2008}. Thermal conduction has been conjectured to compensate
for the cooling losses in the centers of clusters~\citep{Zakamska2003,
  Voit2014NN}, but it is unclear if the actual amount of heat flow can
be as high as expected from traditional theoretical estimates. For
example, mirror instabilities and oblique whistler modes can
potentially suppress electron transport \citep{Komarov2016,
  Riquelme2016, RC2016}. However, the effective volume filling
factor of these processes has not been studied. It is thus still
unclear whether a corresponding suppression of the electron transport
reduces the classical value of the conductivity significantly,
especially in the presence of other mobile anisotropic particle
distributions such as cosmic rays.

Recent simulations~\citep{Ruszkowski2011, Yang2016} have shown that
thermal conduction alone is not strong enough to offset the cooling
losses even if a full Spitzer conduction coefficient along magnetic
field lines is assumed. It may, however, provide part of the heating,
reducing the burden on the blackhole~\citep{Yang2016}. It has also
been found to enhance the mixing of the thermal plasma
in the presence of external sources of turbulence like cosmic ray
driven instabilities~\citep{Sharma2009, Banerjee2014}.

In this Letter, we discuss high-resolution simulations of the formation
of a galaxy cluster, with and without anisotropic thermal conduction.
We investigate the interaction between AGN feedback, magnetic fields
and anisotropic thermal conduction on both the integrated and small
scale properties of the cluster. Our methodology is introduced in
Section~\ref{sec:methods}, the main results are presented in
Section~\ref{sec:results} and interpreted in
Section~\ref{sec:discussion}, and finally, our conclusions are given
in Section~\ref{sec:conc}.
 
 \begin{figure}
\centering
\includegraphics[width=0.49\textwidth]{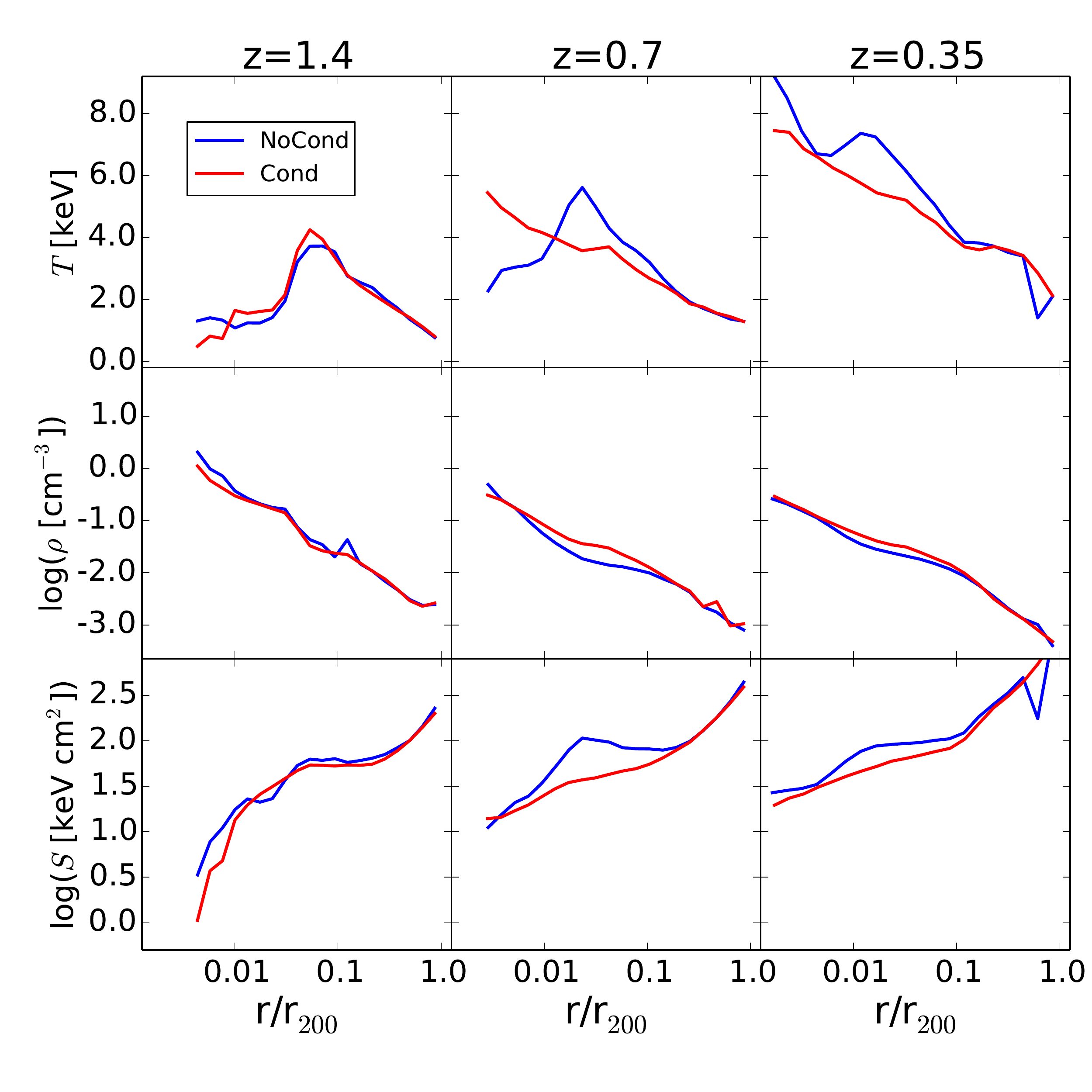}
\caption{Temperature ($T$; top row), density ($\rho$; middle row) and entropy ($S$; bottom row) profiles of the simulated cluster at three representative redshifts, $z=1.4$ (left column), $z=0.7$ (middle column) and $z=0.35$ (right column) in the NoCond (blue curves) and Cond (red curves) runs.}
\label{fig:profile}
\end{figure}

\section{Simulations}
\label{sec:methods}

We have carried out zoom-in cosmological simulations of a massive
($M_{{200}} \sim 6.5 \times 10^{14} \ \rm{M_{\odot}}$) galaxy
cluster as part of the {\tt AESTUS} project (Kannan et al. in
prep). The initial conditions for this cluster were generated from the
Millennium XXL simulation~\citep{Angulo2012} and then rescaled to the
latest Wilkinson Microwave Anisotropy Probe (WMAP)-9
measurements~\citep{Hinshaw2013}: $\Omega_m = 0.2726$,
$\Omega_\Lambda = 0.7274$, $\Omega_b = 0.0456$, $\sigma_8 = 0.809$,
$n_s = 0.963$, and $H_0 = 100\, h\, \rm{km\, s^{-1} Mpc^{-1}} $ with
$h = 0.704$. The high resolution dark matter (DM) and gas masses are
$6.8 \times 10^7 \ \rm{M_{\odot}}$ and
$1.1 \times 10^7 \ \rm{M_{\odot}}$, respectively, with a softening
length of $1.4$ kpc for both particle types. Our mass resolution is
$\sim 1000$ times and our spatial resolution $\sim 30$ times better than
previous simulations attempting to model anisotropic thermal
conduction in a cosmological context~\citep{Ruszkowski2011}. We also
achieve better resolution than idealized non-cosmological simulations
with thermal conduction~\citep{Ruszkowski2010, Parrish2012, Yang2016}
and recent cosmological pure hydrodynamic simulations of
clusters~\citep{Hahn2015, Rasia2015}.

The simulations were performed with the moving-mesh code {\tt
  AREPO}~\citep{Springel2010}, using a module for ideal
magnetohydrodynamics (MHD)~\citep{Pakmor2013}. The simulations employ a
galaxy formation physics model originally developed for the {\tt
  ILLUSTRIS} simulation suite~\citep{Vogelsberger2012, Vogelsberger2013,
  Vogelsberger2014, Vogelsberger2014N}, updated with a new AGN feedback scheme
\citep[][]{Weinberger2016} and modifications to the stellar wind
scheme~(Pillepich et al., in prep).

One of the runs (Cond) additionally includes anisotropic thermal
conduction using the newly developed numerical approach introduced
in~\citet{Kannan2016}. The value of the conduction coefficient is set
to the canonical Spitzer value~\citep{Spitzer1962} along the magnetic
field, with a maximum value of the diffusivity
($\chi \sim \kappa/ C_v \rho$) set to
$5 \times 10^{31} \ \rm{cm^2/s}$~\citep{Ruszkowski2011, Yang2016} and
zero in the perpendicular direction. The conduction routine is not
active for star forming gas cells that follow an equation of state
model for the star-forming interstellar medium~\citep{Springel2003}.
The run without conduction is called NoCond in the following.

\begin{figure}
\centering
\includegraphics[width=0.49\textwidth]{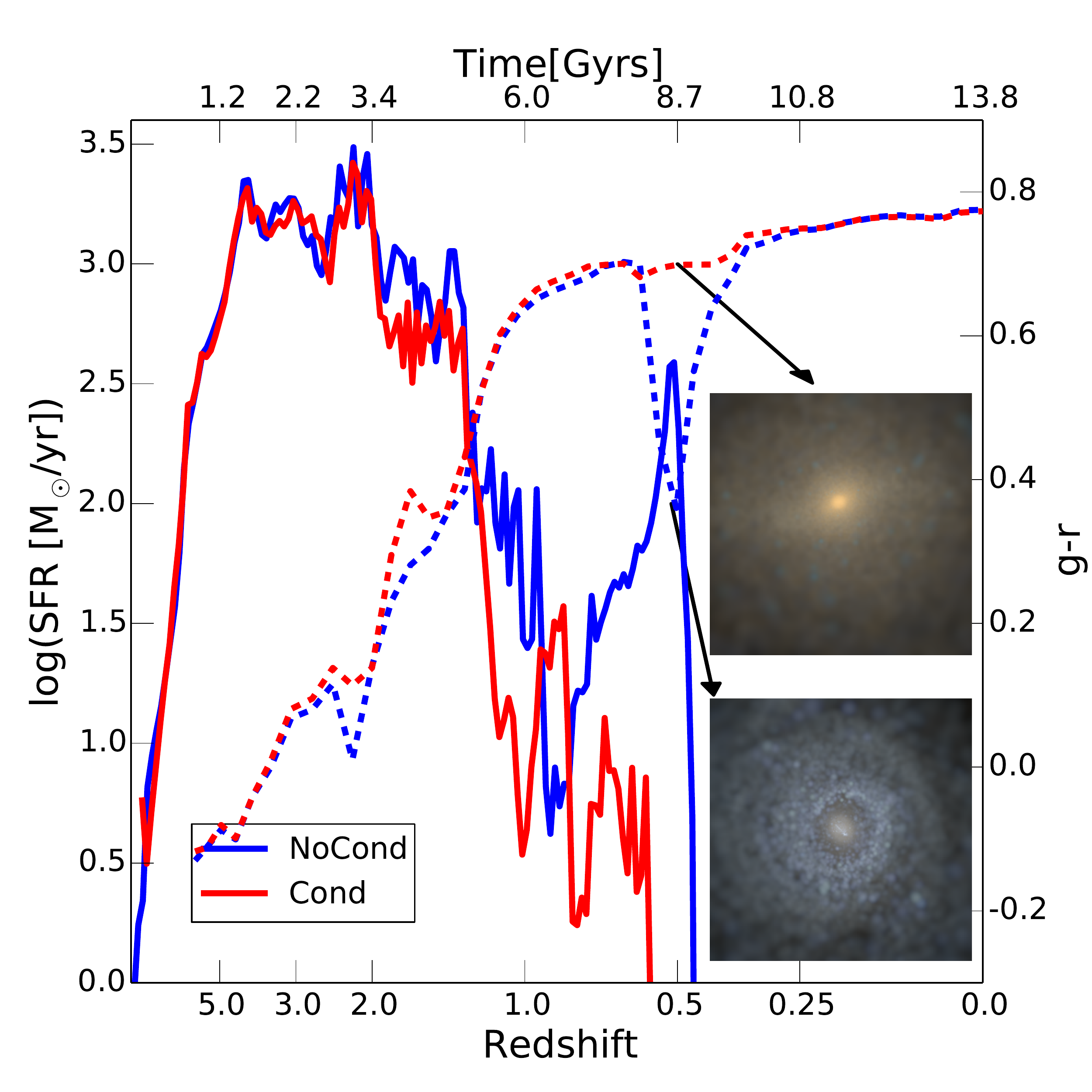}
\caption{SFR (solid curves) and $g-r$ colors of the central galaxy
  (dashed curves) of the simulated cluster as a function of time for
  both the NoCond (blue curves) and Cond (red curves) runs. The insets
  show the synthetic SDSS g-, r-, and i-band composite images of the
  central galaxy of the cluster at $z=0.5$. }
\label{fig:sfr}
\end{figure}

\section{Results}
\label{sec:results}

Fig.~\ref{fig:profile} shows the temperature (top row), density
(middle row) and entropy (bottom row) profiles for both the NoCond
(blue curves) and Cond (red curves) runs at three redshifts. At
$z=1.4$, both simulations exhibit a classic cool-core (CC) structure,
which is characterized by a temperature and entropy drop in the center
and a central high density peak~\citep{Vikhlinin2006,
  Pratt2010}. However, by $z=0.7$, the Cond run has transitioned to a
non cool-core (NCC) cluster state, while the NoCond run still exhibits
a CC structure. By $z=0.35$, both the NoCond and Cond simulations show
a NCC structure.

Another key difference is the lowering of the star formation rate
(SFR) in the Cond run (Fig.~\ref{fig:sfr}) at low redshifts. Above
$z\sim 2$, the SFRs (solid curves) of both the NoCond (blue curves)
and Cond runs (red curves) are similar. After $z\sim 1.4$, when the
transition from CC to NCC happens in the Cond run, a corresponding
decrease in the SFR by almost a factor of three is seen.

 \begin{figure*}
\centering
\includegraphics[width=0.99\textwidth]{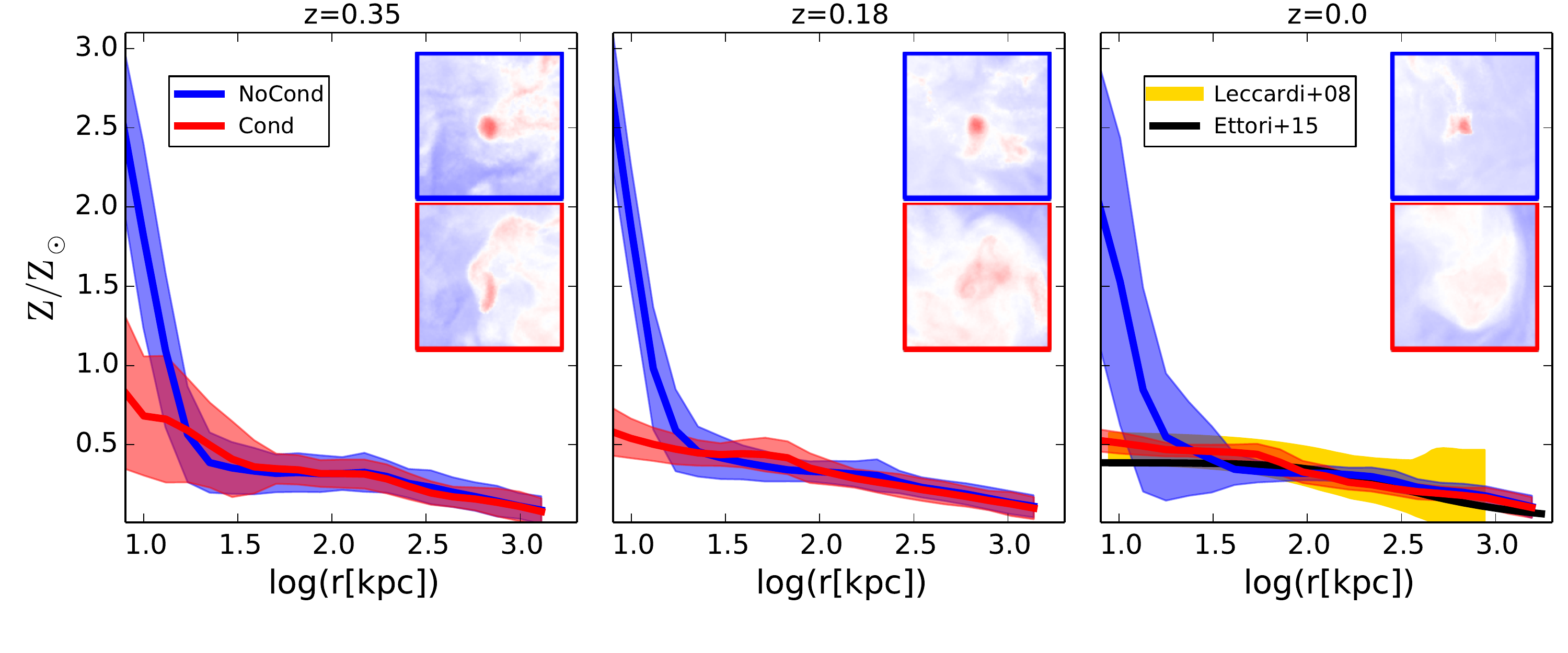}
\caption{Gas metallicity profiles in the NoCond (red curves) and Cond (blue curves) runs at $z=0.35$ (left column), $z=0.18$ (middle column) and $z=0.0$ (right column). The shaded regions denote the $1\sigma$ deviation from the mean. The insets in each panel show the projected mass-weighted gas metallicity maps in the NoCond (blue border) and Cond (red border) runs. The size of the projection box is $(200 \ \rm{kpc})^3$. The observational metallicity estimates at $z=0$ from \citet{Leccardi2008} (yellow shaded region) and \citet{Ettori2015} (solid black curve) are overplotted. }
\label{fig:gmet}
\end{figure*}

At $z=0.95$, the cluster starts undergoing a major ($\sim 1:1$)
merger.  The infall phase of the merger lasts for about $2$~Gyrs, and
the final coalescence of the central galaxies of the merging clusters
takes place at about $z\sim0.6$.  This merger enhances the SFRs in
both runs, but the amount of merger-induced star formation (SF) is
drastically different in the two runs. The Cond run shows only a
modest post-merger SFR of ${\sim 10 \ \rm{M}_\odot/\rm{yr}}$, while
the NoCond run has SFRs that are as high as
${300 \ \rm{M}_\odot/\rm{yr}}$.  Moreover, SF in the Cond run is
completely quenched $\sim 0.5$ Gyrs before the NoCond run.  We note
that although the late time SFRs are very different in the two runs,
the reduction in the total stellar mass in the Cond run is only about
$10\%$.  Furthermore, the stellar metallicities and the stellar ages
of the central galaxy of the cluster are very similar in both runs.
This is because most of the stellar mass has been built
up before $z\sim2$, where the SFRs are generally comparable since
thermal conduction is not very effective at these times.

The discrepancy in the SFRs is also reflected in the intrinsic colors
of the central galaxy (dashed curves and insets in
Fig.~\ref{fig:sfr}), especially in the post merger phase of the
cluster evolution. The ${\rm{g}-\rm{r}}$ color of the central galaxy
in the Cond run is as high as $\sim 0.8$, which places the galaxy in
the red cloud, whereas the high SFRs in the NoCond run reduces the
$\rm{g}-\rm{r}$ color to $\sim 0.4$ for about $2$ Gyrs after the
merger.  Subsequently, AGN feedback turns the galaxy red again.

The right panel of Fig.~\ref{fig:gmet} shows the gas metallicity
profiles of the simulated clusters in the NoCond (blue curve) and Cond
(red curve) runs at $z=0$. The corresponding shaded regions denote the
$1\sigma$ deviation from the mean metallicity.  While the gas phase
metallicities in the outer parts of the cluster are quite similar in
both runs, the difference between them (in both the mean value and
dispersion) within the core of the cluster ($r\leq 100 \ \rm{kpc}$) is
quite striking. The Cond run reproduces observational estimates
(\citealt{Leccardi2008}; gold shaded region $\&$ \citealt{Ettori2015};
solid black curve) of the metallicity profiles in clusters more
faithfully.

The lower SFRs at $z<1$ in the Cond run can in principle explain the
low metallicity values in the center. However, this does not explain
the lower dispersion of metallicities at a fixed radius. In order to
understand this behavior, we plot mass-weighted gas metallicity maps
and the corresponding profiles (Fig.~\ref{fig:gmet}) for both the
NoCond and Cond runs. We chose $z=0.35$ as our starting redshift
because both runs have at that point a quenched galaxy in the center,
and they have entered a relatively quiescent phase of evolution. There
is no metal enrichment due to star formation and the only outflow
mechanism is AGN driven winds. At $z=0.35$, both the runs start out
with a central metallicity core. The core in the NoCond run has a
higher metallicity and is more concentrated because of the larger star
formation rates at late times.  However, by $z=0$, the core in the
Cond run is completely mixed, making the metallicity profile extremely
flat and lowering its dispersion, while the core in the NoCond run
still exists. These results point to the fact that conduction leads to
significantly increased metal mixing, driven by turbulence injected by
the central AGN.

\section{Discussion}
\label{sec:discussion}

\begin{figure}
\centering
\includegraphics[width=0.49\textwidth]{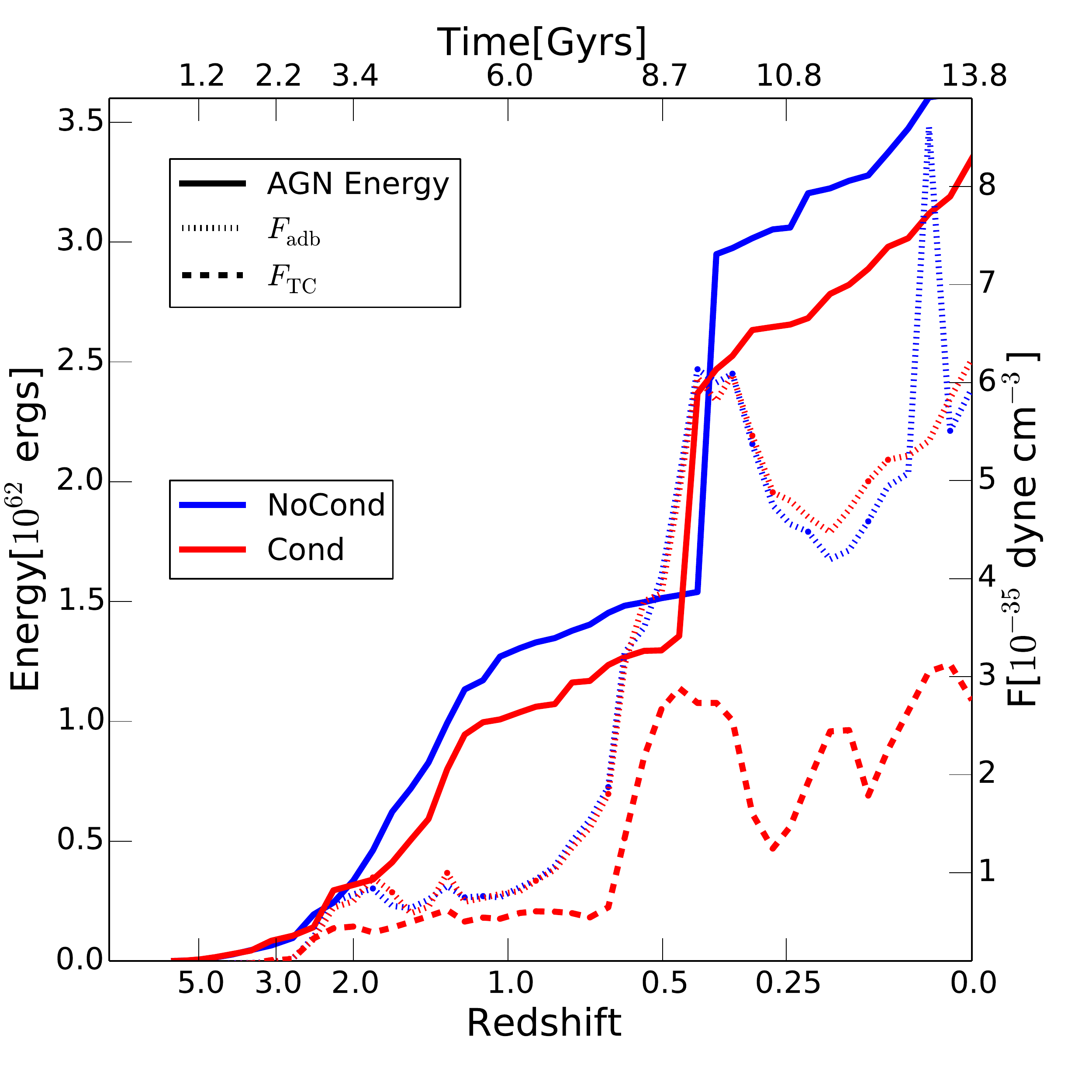}	
\caption{Cumulative amount of AGN energy injected into the ICM by
  the central blackhole (solid curves) in both the quasar and radio
  mode as a function of time. The mean value (within the central $100$
  kpc) of $F_{\rm{adb}}$ (dotted curves) and $F_{\rm{TC}}$ (dashed
  curve) is also plotted as a function of time. The blue curves denote
  the values obtained from the NoCond run, while the red curves show
  the values for the Cond run.}
\label{fig:bhenergy}
\end{figure}

We conclude that the inclusion of anisotropic thermal conduction has a
strong effect on the properties of the ICM (i.e., temperature,
entropy, density, metallicity profiles) and on the characteristics of
the central galaxy (star formation rates, colors etc.). There are
three possible mechanisms through which conduction can cause these
changes. (1) Thermal conduction might force the AGN to inject more
energy into the cluster core by conducting heat outwards, (2)
conductive heating during the CC phase might offset cooling loses, or
(3) thermal conduction might couple the injected AGN energy more
efficiently with the ICM. Fig.~\ref{fig:bhenergy} shows that the
amount of energy deposited by the central AGN in the Cond run (solid
red curve) is consistently lower than in the NoCond run (solid blue
curve) by about $20 - 30\%$. This rules out the first mechanism.

Conductive heating in the CC phase of cluster evolution can in
principle offset the cooling losses in the core, thereby reducing SF.
However, this does not explain the suppression of SF when the Cond run
shows a NCC structure.  More importantly, throughout the CC phase of
cluster evolution the conduction luminosity
($L_{\rm cond} \sim -\kappa \partial{T} / \partial{r}$) in the Cond
run is an order of magnitude lower than both the injected AGN
luminosity and the total cooling luminosity within the cluster core.
Consequently, we can safely conclude that conductive heating cannot be
the full explanation for the observed differences between the two
runs. It can provide at best a part of the energy needed in the CC
phase.

Thus, a more efficient coupling of the injected AGN feedback energy
with the surrounding ICM, mediated by conduction, seems to be the most
plausible explanation. The AGN feedback model used in our simulations
distinguishes between a high accretion rate quasar-mode feedback
channel, modeled through local thermal energy injection, and a low
accretion rate kinetic feedback mode, imparting momentum into the
surrounding gas~\citep{Weinberger2016}.  The direction of the momentum
injection is stochastic such that on average it is isotropic.  The
quasar mode feedback dominates at high redshifts, while the kinetic
feedback mechanism becomes important below $z\sim1.5$.

\begin{figure}
\centering
\includegraphics[width=0.49\textwidth]{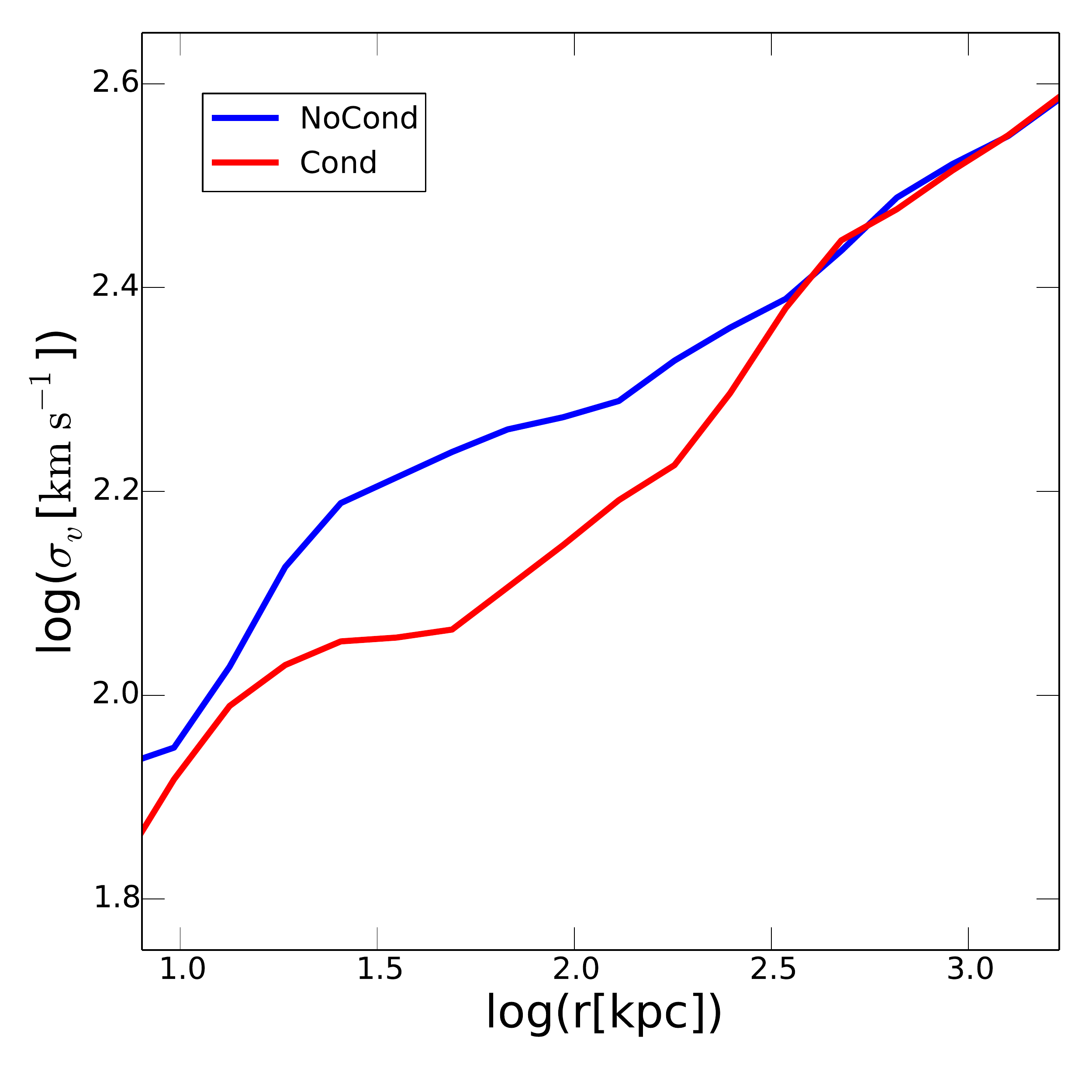}
\caption{Average (between $z=0.35$ and $z=0$) one-dimensional
  velocity dispersion profiles in the NoCond (blue curve) and Cond
  (red curve) runs.}
\label{fig:disp}
\end{figure}

The metallicity evolution between $z=0.35$ and $z=0.0$ clearly
demonstrates that there is more efficient turbulent mixing of the
thermal plasma (and consequently metals) within the cluster core which
can in principle explain the increased coupling efficiency between the
AGN feedback energy and the ICM.  However, the average (between
$z=0.35$ and $z=0.0$) one-dimensional velocity dispersion profiles of
the ICM (Fig.~\ref{fig:disp}) in the NoCond (blue curve) run
($150 \ \rm{km \ s^{-1}}$) is higher than in the Cond (red curve) run
($100 \ \rm{km \ s^{-1}}$), especially in the cluster core. This seems
to suggest that there is more metal/plasma mixing in-spite of lower
turbulent velocities in the Cond run.

The ability of external turbulence to efficiently mix a stratified
plasma depends on its convective stability, which, for a pure
hydrodynamic plasma, is decided by its entropy gradient. Specifically,
the plasma is stable as long as $\partial{S}/\partial{r}>0$
\citep[Schwarzschild criterion,][]{SCbook}. This is generally true for
all observed clusters \citep[eg.][]{Vikhlinin2006} and hence the ICM
is conventionally thought to be convectively stable. In the presence
of external turbulence, a fluid element in a stably stratified
atmosphere which is adiabatically displaced from its equilibrium
position by a small amount $\delta r$ will experience a buoyant
restoring force
$F_{\rm{adb}} \sim \rho g ({\rm d}\, {\rm ln}S / {\rm d}r)\delta r$
\citep{Ruszkowski2010}, causing oscillations around its equilibrium
position at the classical Brunt-V\"ais\"al\"a frequency. If the
turbulent driving force ($F_{\rm{turb}}$) is larger than the buoyant
restoring force ($F_{\rm{adb}}$) then it can induce mixing.

This picture changes in the presence of anisotropic thermal conduction
because it fundamentally changes the response of the plasma to
perturbations. Provided ${\rm d}T/{\rm d}r \ne 0$, conduction along
magnetic field lines causes the ICM to be (formally) buoyantly
unstable regardless of the temperature and entropy gradients. When
${\rm d}T/{\rm d}r > 0$ (for CCs in the central cooling region) the
ICM is unstable to the heat-flux-driven buoyancy
instability~\citep[HBI]{Quataert2008}, and when
${\rm d}T/{\rm d}r < 0$ (for all clusters on large scales) it is
unstable to the magneto-thermal
instability~\citep[MTI]{Balbus2000}. As a consequence, any amount of
external turbulence will instantly mix the already convectively
unstable (i.e., zero restoring force) thermal plasma without an energy
penalty \citep{Sharma2009}.  Even in the saturated state of these
instabilities the buoyant restoring force is proportional to the
temperature gradient
($F_{\rm{TC}} \sim \rho g ({\rm d}\, {\rm ln}T / {\rm d}r) \delta r $)
instead of the entropy gradient \citep{Sharma2009}.

Our blackhole feedback model is characterized by self-regulation,
i.e.~it keeps injecting energy until the cooling losses are accounted
for. This implies that the AGN will essentially supply higher
and higher turbulent energy until the injected turbulent driving force
is larger than the buoyant restoring force, at which point it will
induce mixing, isotropizing the injected energy and stopping cooling
in the cluster core. The higher the buoyant restoring force, the
larger the amount of turbulence injected by the AGN in order to
induce mixing.  

Fig.~\ref{fig:bhenergy} shows that the restoring forces in the NoCond
and Cond runs, calculated using the entropy gradient ($F_{\rm{adb}}$,
blue and red dotted curves, respectively), are quite similar. When we
account for the fact that the buoyant response of an anisotropically
conducting plasma is fundamentally different from that of a pure
hydrodynamic fluid and calculate the restoring force using the
temperature gradient ($F_{\rm{TC}}$, red dashed line), we find
lower restoring forces in the Cond run, which essentially explains the
lower turbulent velocities and increased rate of plasma mixing in this
run.

We note that $F_{\rm{TC}}$ is in principle only valid in the saturated
state of the instabilities, otherwise the restoring forces are
essentially zero. Hence, $F_{\rm{TC}}$ represents an upper limit to
the restoring force in an anisotropically conducting plasma.  It is
very difficult to assess the state of these instabilities in fully
cosmological simulations. We can however get some insights by noting
that the timescale between successive AGN bursts,
$\sim 10 \ \rm{Myrs}$, is much smaller than the growth timescales for
these instabilities, which is of the order of $\sim 500 \ \rm{Myrs}$
\citep{Parrish2012}. This implies that the HBI and MTI instabilities
are unlikely to saturate or even grow considerably between successive
AGN injection events. Besides, Fig.~\ref{fig:disp} clearly shows that
it is not the additional turbulence generated by these instabilities
that causes the plasma to mix, rather, they change the buoyant
response of the ICM and make it more prone to mixing. It is the
injection of turbulence into this modified state that causes the
efficient mixing of the plasma. Therefore, the AGN kinetic wind power
needed to induce gas mixing in the core is reduced. This allows the
more efficient utilization of the injected energy in the Cond run,
thereby drastically improving the coupling between the injected
feedback energy and the ICM. The enhanced mixing of gas of different
entropies will also contribute to a flattening of the entropy profiles
in cool core phases of evolution.

Moreover, the increased mixing in the Cond run randomizes the magnetic
field orientation, which in turn isotropizes the direction of conductive
heat flow. This is particularly effective in redistributing the energy
in the quasar mode of the AGN feedback, and in the high
temperature regions \citep{Weinberger2016} formed when the AGN
kinetic winds shock against the ICM.  Therefore, it seems that
conduction enables turbulence, and turbulence enables conduction
\citep{Sharma2009, Ruszkowski2010}. Although we have shown that there
is more mixing in the Cond run, we have not rigorously quantified this
mechanism. This is beyond the scope of the current study and is left
for future work.

\section{Conclusions}
\label{sec:conc}

We have presented cosmological MHD simulations of the formation of a
galaxy cluster, comparing calculations with and without anisotropic
thermal conduction. These are the first simulations to
self-consistently include and quantify the effect of thermal
conduction on both the integrated and small-scale properties of galaxy
clusters. Our main results can be summarized as follows:
\begin{itemize}
\item Thermal conduction causes an earlier disruption of the cool
  core, and a subsequent reduction of the star formation rates by more
  than an order of magnitude at low redshift. The central gas phase
  metallicity gradients and dispersions are also reduced, despite an
  overall lower amount of AGN feedback energy injected into the ICM.

\item The coupling between the AGN feedback energy and the ICM is
  effectively enhanced in the presence of anisotropic thermal
  conduction. It is considerably easier to mix thermal plasma in the
  presence of conduction, because the plasma is unstable irrespective
  of the temperature or entropy gradient and thus already prone to
  mixing. The restoring buoyancy forces are reduced, leading to
  efficient mixing even with low levels of external turbulent
  driving. This helps to isotropize the injected AGN feedback energy,
  thereby quenching the clusters more efficiently.
\end{itemize}

We have also simulated two other, less massive clusters
($M_{\rm halo} \sim 2 \times 10^{14}\, {\rm M}_\odot$ and
$ 6 \times 10^{14}\, {\rm M}_\odot $) at lower resolution. The general
trends (earlier termination of the cool core, lower SFRs, and lower central
metallicities and dispersion etc.) are very similar for these
clusters, suggesting that our conclusions robustly apply to other
systems as well.
 
We stress that although thermal conduction helps in quenching star
formation, it is not sufficient on its own for stabilizing clusters or
converting a cool core to a non cool core (at least not for these cluster masses). It
only amplifies the effect of external turbulent driving. The external
source of turbulence can in principle take many forms, such as
mergers, cosmic ray driven convection, etc., and is not limited to the
kinetic AGN winds examined here. We thus expect that the
importance of anisotropic thermal conduction carries over to other
forms of feedback as well.

\begin{acknowledgements}

We thank Eliot Quataert for useful comments and discussions. MV acknowledges support through an MIT RSC
award and the Alfred P. Sloan Foundation. RW, VS and RP acknowledge
support by the European Research Council under ERC-StG grant EXAGAL
308037. CP acknowledges support by the European Research Council under
ERC-CoG grant CRAGSMAN-646955. CP, RW, VS and RP also acknowledge
support from the Klaus Tschira Foundation.

\end{acknowledgements}

\end{document}